\documentclass[10pt,letterpaper]{article}
\usepackage[top=0.85in,footskip=0.75in,marginparwidth=2in]{geometry}

\usepackage[utf8]{inputenc}

\usepackage{cite}

\usepackage{nameref,hyperref}

\usepackage[right]{lineno}

\usepackage{microtype}
\DisableLigatures[f]{encoding = *, family = * }

\raggedright
\usepackage{changepage}

\usepackage[aboveskip=1pt,labelfont=bf,labelsep=period,singlelinecheck=off]{caption}

\usepackage{mathtools}
\usepackage{bm}
\usepackage{amssymb}

\makeatletter
\renewcommand{\@biblabel}[1]{\quad#1.}
\makeatother

\usepackage{lastpage,fancyhdr,graphicx}
\usepackage{epstopdf}
\fancyheadoffset[L]{0.25in}
\fancyfootoffset[L]{0.25in}

\usepackage{color}

\definecolor{Gray}{gray}{.25}

\usepackage{graphicx}
\usepackage{ragged2e}

\begin{document}
\vspace*{0.35in}

\begin{flushleft}
{\Large
\textbf\newline{Multi-phase Deformable Registration for Time-dependent Abdominal Organ Variations}
}
\newline
\\
Seyoun Park\textsuperscript{a,*},
Elliot K. Fishman\textsuperscript{a},
Alan L. Yuille\textsuperscript{b,c},
\\
\bigskip
\textsuperscript{c} Department of Radiology and Radiological Science, Johns Hopkins University, USA
\\
\textsuperscript{a} Department of Computer Science, Johns Hopkins University, USA
\\
\textsuperscript{d} Department of Cognitive Science, Johns Hopkins University, USA
\\
\bigskip
*seyoun.park@gmail.com

\end{flushleft}

\justify
\section*{Abstract}

Human body is a complex dynamic system composed of various sub-dynamic parts. Especially, thoracic and abdominal organs have complex internal shape variations with different frequencies by various reasons such as respiration with fast motion and peristalsis with slower motion. CT protocols for abdominal lesions are multi-phase scans for various tumor detection to use different vascular contrast, however, they are not aligned well enough to visually check the same area. In this paper, we propose a time-efficient and accurate deformable registration algorithm for multi-phase CT scans considering abdominal organ motions, which can be applied for differentiable or non-differentiable motions of abdominal organs. Experimental results shows the registration accuracy as $0.85 \pm 0.45mm$ (mean $\pm$ STD) for pancreas within 1 minute for the whole abdominal region.


\section{Introduction}
\label{sec:Introduction}

Human body is a complex dynamic system composed of various sub-dynamic parts. Especially, thoracic and abdominal organs have complex internal shape variations with different frequencies by various reasons. In short time, the major large motion is induced by respiration, but there are other factors such as peristalsis, e.g. stomach expansion or shrinkage as well according to scanning protocols. In longer terms, pose changes and anatomical changes such as weight loss and tumor growth can happen as well. These variations make difficulties to effectively interpret and computationally use medical images acquired at different time.

Internal motion does not only cause shape changes, but also can bring density variation, which can cause even in the same modalities. In addition, according to scan protocols, internal intensities can be shown as different values according to vascular structures in case of contrast enhanced images as shown in Figure 1.

During last two decades, deformable image registration (DIR) is one of the most extensive topics in medical image analysis. Various surveys \cite{Maintz1998,Sotiras2013,Viergever2016} explains how many studies have been done. It can be categorized from different aspects, such as image modalities, dimensions, intra-/inter-subjects, target regions(e.g. brain, head and neck, chest, abdomen, etc.), and target applications. Despite of the long history and extensive works of DIR on medical images, it suffers from various factors, especially for the abdominal regions which has complex anatomical structures with low-contrast between soft tissues, and continuous motion with various frequencies. 

\begin{figure*}
\centering
\includegraphics[width=0.45\textwidth]{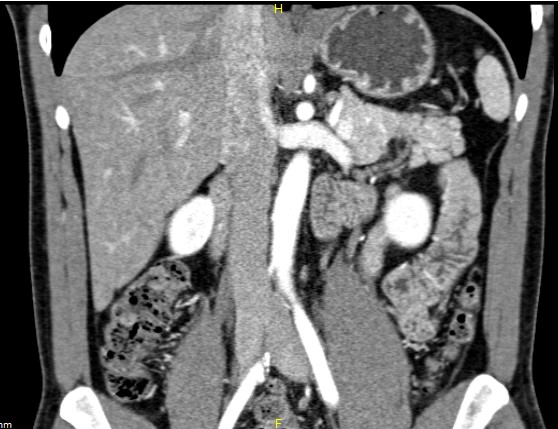}
\includegraphics[width=0.45\textwidth]{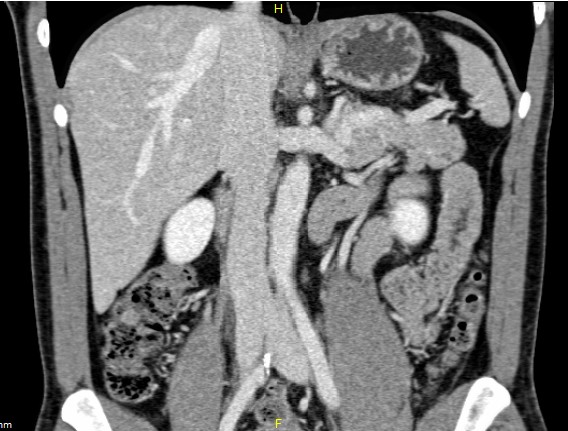}
\label{fig:multiphase}
\caption{An example of two-phase CT scans. Arterial (left) and venous (right) phases.}
\end{figure*}

In this study, we are targeting the intra-subject registration of abdominal regions considering $(1)$ multi-phase intra-subject CT scans with $(2)$ short time motions and $(3)$ possible discontinuities of the motion. 

Recently, learning-based approaches based on convolutional neural networks has been introduced by supervised \cite{Fan2018} or unsupervised ways \cite{Dalca2019}, or by generative models \cite{Krebs2019}. However, in these approaches, deformation fields are implicitly modeled by introducing latent variables. In addition, there are still critical limitations to apply to thin-slice CT volumes in terms of number of dataset and required memory, especially to apply abdominal region. 

To solve this problem, we applied direction-dependent deformable registration \cite{Richberg2012} between multi-phase CT images, especially for abdominal region. As the conventional deformable registration approach, we explicitly model the deformation field. To consider intensity differences between multi-phase scans, we adapted local self structural similarities introduced for multi-modal deformable registration in \cite{Heinrich2012,Reaungamornrat2016} to find matching point by allowing direction-dependent local discontinuities. Different elastic properties computed based on intensities allow different levels of local rigid and full deformable transformations. We implemented our algorithm on efficient communication between CPU and GPU which showed very efficient computation with high fidelity by experiments.

\section{Methods}
\label{sec:Methods}

Given a 3D volume of interest (VOI) of a scanned CT image $\textrm{V} \subset \mathbb{R}^3$, the goal of this study is to find the optimal deformation vector field which minimizes the difference between the target and the deformed source volume. Let us denote the fixed target image on the domain $\textrm{V}$ as $I_{F}$ and the moving source volume, which will be deformably aligned to $I_{F}$, as $I_{M}$, where $I:V \rightarrow \mathbb{R}$. In our problem, we are assuming that $I_{M}$ and $I_{F}$ are scanned from the same CT coordinate within short-time difference for the same subject. Our goal is to find the optimal deformation vector field $\textrm{T}: \mathbb{R}^3 \rightarrow \mathbb{R}^3$, which minimizes the structural differences between $I_{F}$ and the deformed volume $(I_{M} \circ \textrm{T})$. 

\subsection{Direction-dependent regularization}

As we are considering possible discontiunities of motiong such as slipping, $T$ is not a differentiable map and the diffeomorphism is not proper to be directly applied to our problem. Therefore, we are defining the registration problem to find the optimal $\textrm{T}$ which minimizes:

\begin{equation}\label{eq:Energy}
  E(\textrm{T}) = \textrm{D}({I_{F} , I_{M} \circ \textrm{T}}) + \alpha R(\textrm{T}),
\end{equation}
where $\textrm{D}(I_{1} , I_{2})$ represents the difference between two images, $\alpha$ is coefficient, and $R(T)$ is the regularization term of the deformation. $\textrm{T}$ is the function $\textrm{T}(\textbf{x}) = \textbf{u}$ of the motion vector $\textbf{u}$ at a voxel position $\textbf{x}$.  To apply potential discontinuities between organs at the boundary, we divide the organ motion $\{\textbf{u}\}$ in two directional component, $\textbf{u}^{\perp}$ and $\textbf{u}^{||}$, which are perpendicular and parallel to the organ boundary, respectively, as proposed in \cite{Richberg2012}, The regularization term is then can be modeled by direction-dependent terms as

\begin{equation}\label{eq:Reg2}
\begin{split}
  R(T)= {1 \over 2 } \sum_{l=1}^{l=3} 
  & { \int_\Omega {\omega ||\nabla u_l ^{\perp} ||^2+ (1-\omega)||\nabla u_l ||^2 dx} +} \\
  & {\int_{\Gamma} {\omega||\nabla u_l || ^2 dx} + \int_{\Omega / \Gamma} { \omega || \nabla u_l ^{||} ||^2 dx }} ,
\end{split}
\end{equation}
where $\omega$ is a bell shaped weight function. 
The solution of (\ref{eq:Energy}) can be solved in variational approach. By Euler-Lagrange equation is going to be (\ref{eq:ELSol}) \cite{Schmidt-Richberg2009,Richberg2012} 

\begin{equation}\label{eq:ELSol}
  \textbf{f} ( \textbf{u} ) - \alpha \textbf{c} _{R} (\textbf{u} ) = 0,
\end{equation}
by introducing the force term $\textbf{f}(\textbf{u})$ and the correction term $\textbf{c}_{R}(\textbf{u})$. For the regularizer in \ref{eq:Reg2}, $\textbf{c}_{R}(\textbf{u})$ becomes 

\begin{equation}
\label{eq:correction}
\textbf{c}_{R} (\textbf{u}) = \nabla \omega \nabla \textbf{u}^{\perp} + \nabla(1-\omega)\nabla \textbf{u} + \overline{\nabla}\omega \overline{\nabla}\textbf{u}^{||},
\end{equation}
where $\overline{\nabla}$ is the gradient only computed from inside or outside of target organs not edges. 
One of the way to solve the \ref{eq:ELSol} is to apply iterative scheme as \cite{Richberg2012}.

\begin{equation}\label{eq:updateU}
  \textbf{u} ^{(k+1)} = \textbf{u} ^{k} + \tau \left( {\textbf{f} _D ({\textbf{u} ^{(k)}}) - \alpha \textbf{c}_R ({\textbf{u} ^{k} })}\right),
\end{equation}
where $k$ is the iterative step and $\tau$ is user-defined constant for the step size. 

\subsection{Local self-similarity}

A modality independent neighborhood descriptor (MIND) \cite{Heinrich2012} builds from the concept of self-similarity which can be used to capture corresponding local structures. It is computed with the configuration of neighboring voxels, called a stencil \cite{Heinrich2012} as: 

\begin{equation}
    \label{MIND}
    \textrm{m}(I, \textbf{x}, \textbf{r}) = {1 \over n} \exp{\left(- { {D_{p}(I, \textbf{x}, \textbf{x}+\textbf{r})} \over {V(I, \textbf{x})} }\right)},
\end{equation}
where $n$ is a normalization factor and $\textbf{r}$ is the search region. The numerator is patch-based distance of the volume and defined as 

\begin{equation}
D_{p}(I, \textbf{x}_{1}, \textbf{x}_{2}) = \sum_{\textbf{p} \in P} {\left(I, \textbf{x}_{1} + \textbf{p}) + I(\textbf{x}_{2}+\textbf{p})\right)}^{2}. 
\end{equation}
The denominator is the estimation of the local variance and computed by the mean of the patch distances within a six-neighborhood $\textbf{n} \in N$.

\begin{equation}
    \label{eq:Variance}
    V(I, \textbf{x}) = {1 \over 6} \sum_{\textbf{n} \in {N}} {D_{p} \left( I, \textbf{x}, \textbf{x} + \textbf{n} \right)}.
\end{equation}
By applying MIND $m$ to compute difference between two images, $\textbf{f}_{D} (\textbf{u})$ at pixel location $\textbf{x}$ in (\ref{eq:ELSol}) becomes similar to demons algorithm: 

\begin{equation}
    \label{eq:Diff}
    \textbf{f}_{D}(\textbf{x}, \textbf{u}) = {{m(I_{M}, \textbf{x}) - m(I_{F}, \textbf{x} - \textbf{u}(\textbf{x}))} \over {||\nabla m(I_{M}, \textbf{x})||^{2} + \kappa^{2}}} \nabla m(I_{M}, \textbf{x}),
\end{equation}
where $\kappa$ is a constant to avoid unstable computation and $0.5$ was used in our experiments.

The iterative algorithm then can be step-wised by computing (\ref{eq:Diff}) for the given $\{\textbf{u}^{k}\}$ at the $k$-th iteration, updating $\{\textbf{u}^{\perp}\}$ and $\{\textbf{u}^{||}\}$, computing (\ref{eq:correction}), and updating $\{\textbf{u}^{(k+1)}\}$ with (\ref{eq:updateU}).

\subsection{Organ-map generation}
This step is not necessary and different estimation can be applied for the initial estimation of boundaries of target organs, we applied semantic multi-organ segmentation \cite{Wang2019} to generate the probability map of the abdominal organs. Figure 2 shows an example of pancreas and liver probability maps in an axial image of a same subject. The directional component can be divided by normal vectors $\{\textbf{n}\}$ of the map as, $\textbf{u}^{\perp} (\textbf{x}) = \left(\textbf{u}(\textbf(x) \cdot \textbf{n}(\textbf{x})\right)\textbf{n}(\textbf{x})$ and $\textbf{u}^{||} (\textbf{x}) = \textbf{n}(\textbf{x}) - \textbf{u}^{\perp}(\textbf{x})$. 

\begin{figure*}
\label{fig:organmap}

\includegraphics[width=0.45\textwidth]{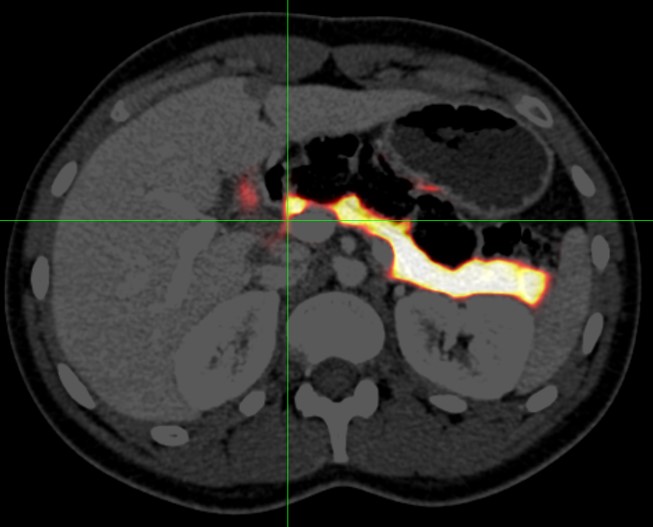}
\includegraphics[width=0.45\textwidth]{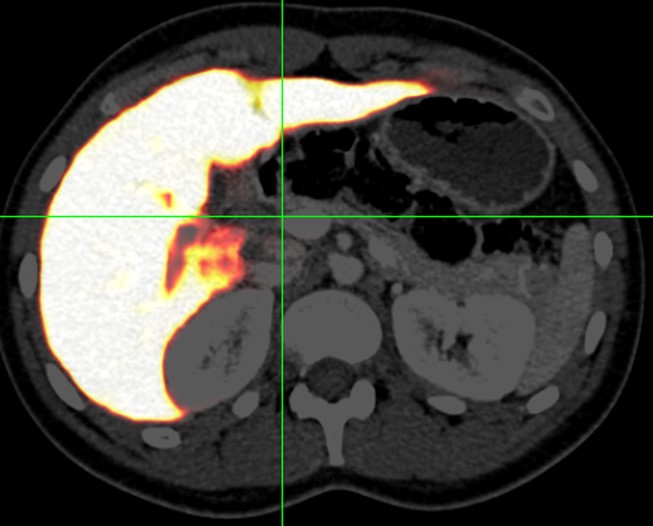}
\includegraphics[width=0.033\textwidth]{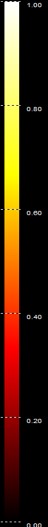}
\caption{Organ probability maps for pancreas (left) and liver (right). The probability maps overlaid on the axial image of CT.}

\end{figure*}

\subsection{Hierarchical modeling}
For the efficient computation, we applied hierarchical approach for the computation. Three levels of down-sampled volumes were generated and self-similarities were computed as a pre-processing, and deformation was computed from the most coarse-level and used as initial value for the finer-level $\{\textbf{u}\}$ vector field.

\section{Experimental Results}

\subsection{Dataset}
\label{sec:Dataset}

We evaluated our methods on $176$ pancreatic ductal adenocarcinoma (PDAC) patients and $94$ renal dornors who can be considered as ``normal'' CT images of normal cases under an IRB (Institutional Review Board) approved protocol. PDAC patients were scanned on a $64$-slice multidetector CT scanner (Sensation $64$, Siemens Healthineers) or a dual-source multidetector CT scanner (FLASH, Siemens Healthineers) and normal controls were scanned on a 64-slice multidetector CT scanner (Sensation 64, Siemens Healthineers). PDAC patients and renal donors were injected with $100-120 mL$ of iohexol (Omnipaque, GE Healthcare) at an injection rate of 4-5 mL/sec. Scan protocols were customized for each patient to minimize dose but were in the order of 120 kVp, effective mAs of $300$, pitch of $0.6-0.8$, and collimation of $64 \times 0.6 mm$ (for $64$-slice scanner) of $128 \times 0.6 mm$ (for dual source scanner). Arterial phase imaging was performed with bolus triggering, usually between 25-30 seconds post-injection and venous phase imaging was performed at 60 seconds. All images were reconstructed into $0.75 mm$ slices with $0.5mm$ increments. CT scans are composed of $(319-1051)$ slices of $(512 \times 512)$ images, and have voxel spatial resolution of $\left([0.523-0.977] \times [0.523-0.977] \times 0.5 \right)mm^{3}$.

Liver and pancreas were segmented in a manual way by trained scientists and checked by experienced radiologists. To obtain consistent set of annotations for the same patient, a labeler annotated for both of arterial and venous phases for the same patient in a consecutive manner. However, to collect algorithm-independent datset, labelers annotated arterial and venous phase without any automatic alignment. To evaluate accuracy of the algorithm, we propagate the manual contours of abdominal organs of source images by applying deformation vector field to the target image and computed the difference from the manual segmentation of the target image. 

\subsection{Implementation}
\label{sec:Implementation}

Our algorithm was written in C++ for CPUs and NVidia CUDA for GPUs. All the computations in each iteration step was parallelized into GPUs and updated to the GPU memory. It was tested on a workstation with Intel i7-6850K CPU, NVidia TITAN X (PASCAL) GPU with 12GB memory.

\subsection{Evaluation}
\label{sec:Evaluation}

The performance was evaluated using 4 different metrics and it includes surface distances (SD, mean $\pm$ standard deviation) and Dice-Sorenson similarity coefficient (DSC) between manual contours of the target organs and propagated contours from the moving images. To compare image similarity, we used structural similarity (SSIM) \cite{Wang2004} and normalized cross-correlation (NCC) between $\textrm{I}_{T}$ and $(\textrm{I}_{S} \circ T)$. We compared our algorithm with popular deformable registration algorithms, diffeomorphic demons (DD) \cite{Vercauteren2009}, MIND elastic \cite{Heinrich2012}, and SyN \cite{Avants2008}. 

\begin{table}
\caption{Performance evaluation of image similarities}\label{tab:image}

\begin{tabular}{|l|c|c|}
\hline
Method &  SSIM & NCC \\
\hline
DD & $0.901$  & $0.948$ \\
MIND & $0.889$  & $0.935$ \\
SyN* & $0.902$ & $0.943$ \\
Ours & $\textbf{0.924}$ & $\textbf{0.966}$ \\
\hline
\end{tabular}
\end{table}

\begin{table}
\caption{Performance evaluation of target structures}\label{tab:structure}

\begin{tabular}{|l|c|c|c|c|}
\hline
Method &  Pancreas SD & Pancreas DSC(\%) & Liver SD & Liver DSC(\%) \\
\hline
DD & $1.39\pm 0.69$ & $87.16$ & $1.35\pm 0.48$ & $93.52$ \\
MIND & $1.53\pm 0.57$ & $86.24$ & $2.52\pm 0.59$ & $93.85$ \\
SyN* & $1.37\pm 0.63$ & $87.32$ & $1.23\pm 0.51$ & $95.32$ \\
Ours & $\textbf{0.85}\pm \textbf{0.45}$ & $\textbf{89.04}$ & $\textbf{0.78} \pm \textbf{0.37}$ & $\textbf{98.52}$ \\
\hline
\end{tabular}
\end{table}

In case of SyN*, not all the cases were computed as it failed for the big volumes or consumed too long time for the computation. For the failure cases, we used one level coarser volume. Therefore, it can't be direct comparison, but we report for the reference. 
As shown in in Tables \ref{tab:image} and \ref{tab:structure}, the proposed approach outperforms other algorithms in terms of image similarities and propagated structural similarities. Computation time for all the cases was less than 1 minute except for organ-map generation with 48.59sec as average. Computation time can be various by various factors such as the maximum number of iterations, however, it is practically acceptable in various domains. 

\begin{figure*}
\label{fig3}
\centering
\includegraphics[width=0.4\textwidth]{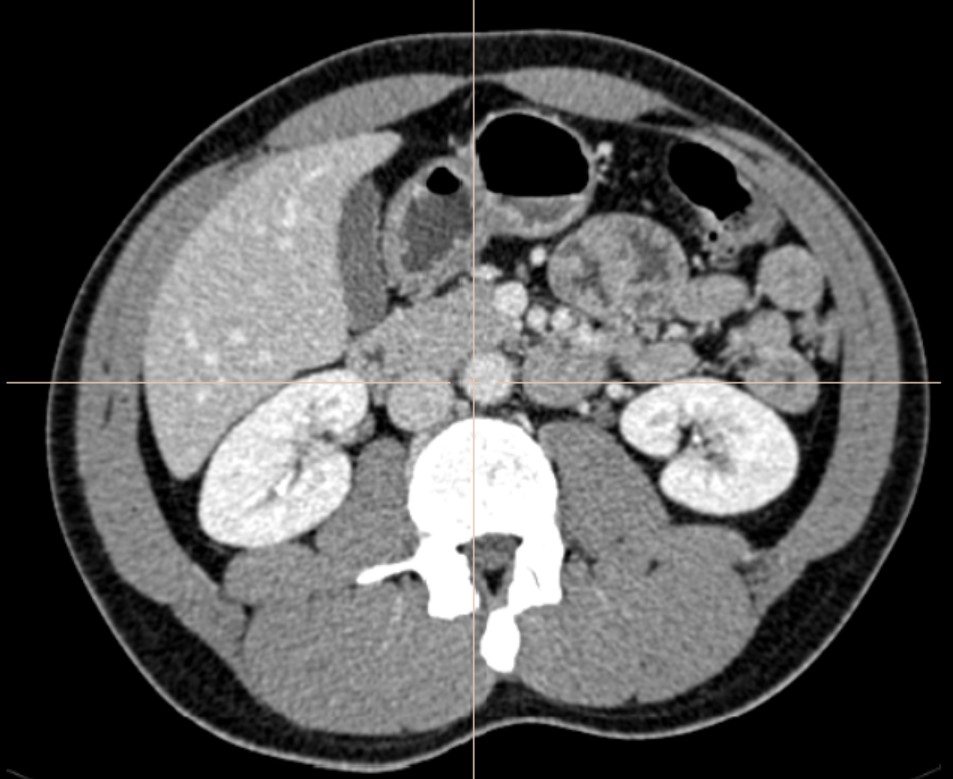}
\includegraphics[width=0.36\textwidth]{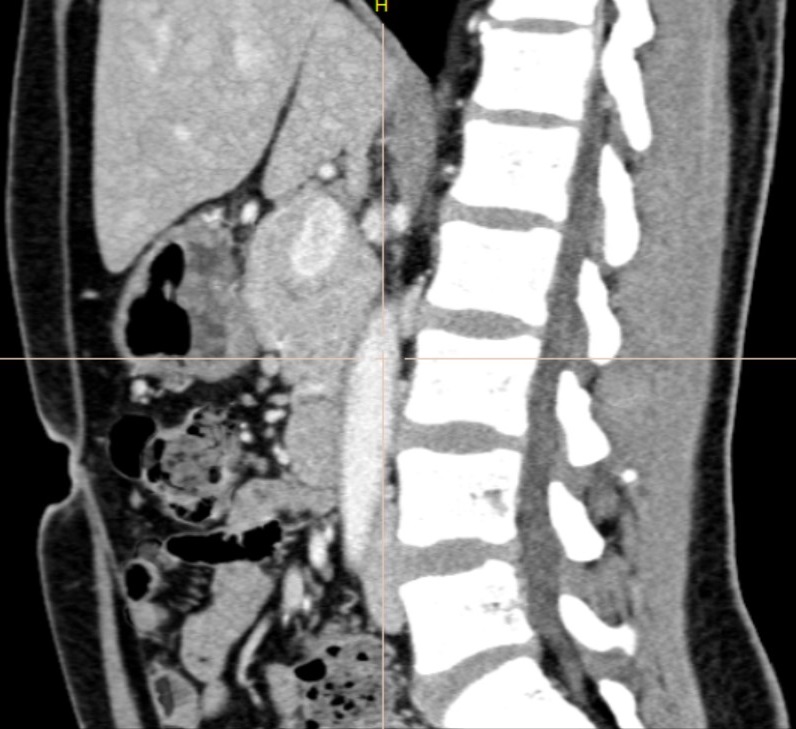}
\includegraphics[width=0.4\textwidth]{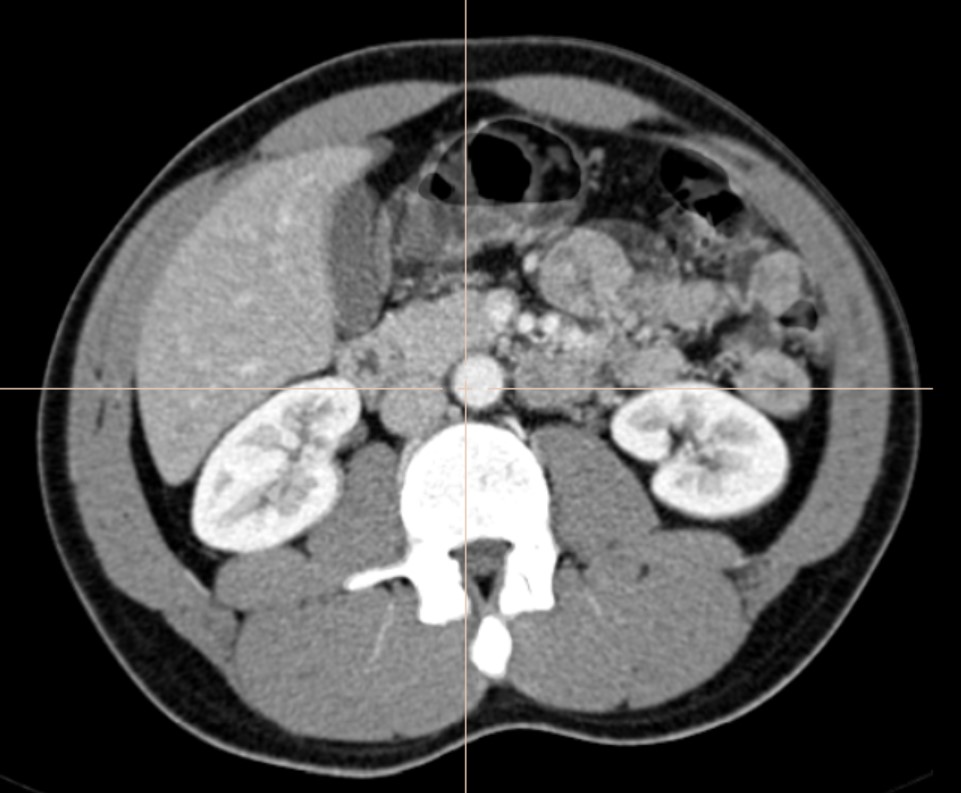}
\includegraphics[width=0.36\textwidth]{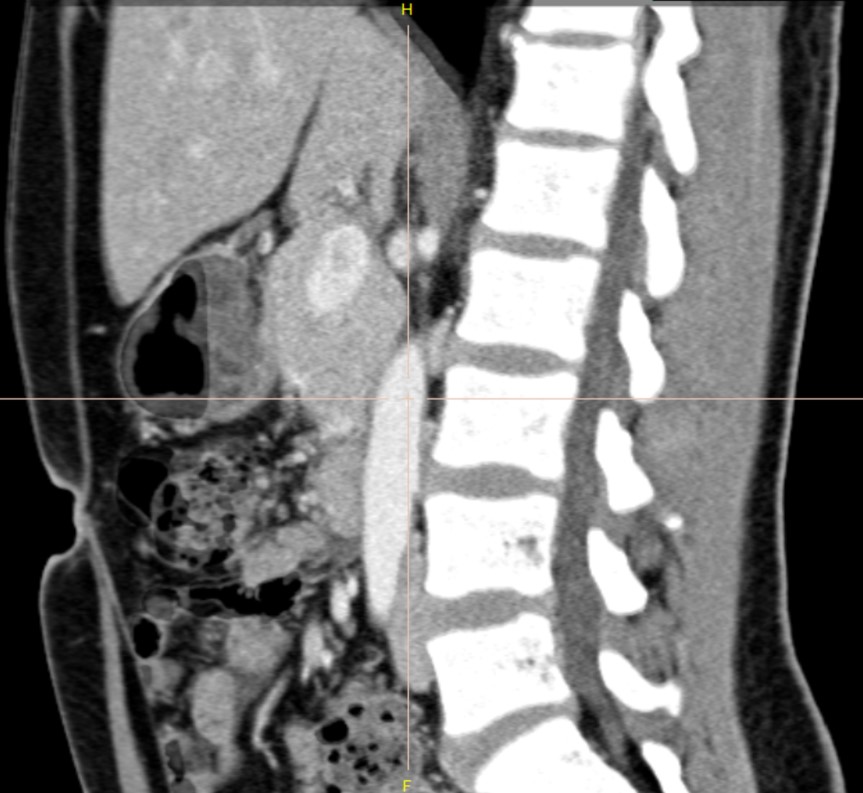}
\includegraphics[width=0.4\textwidth]{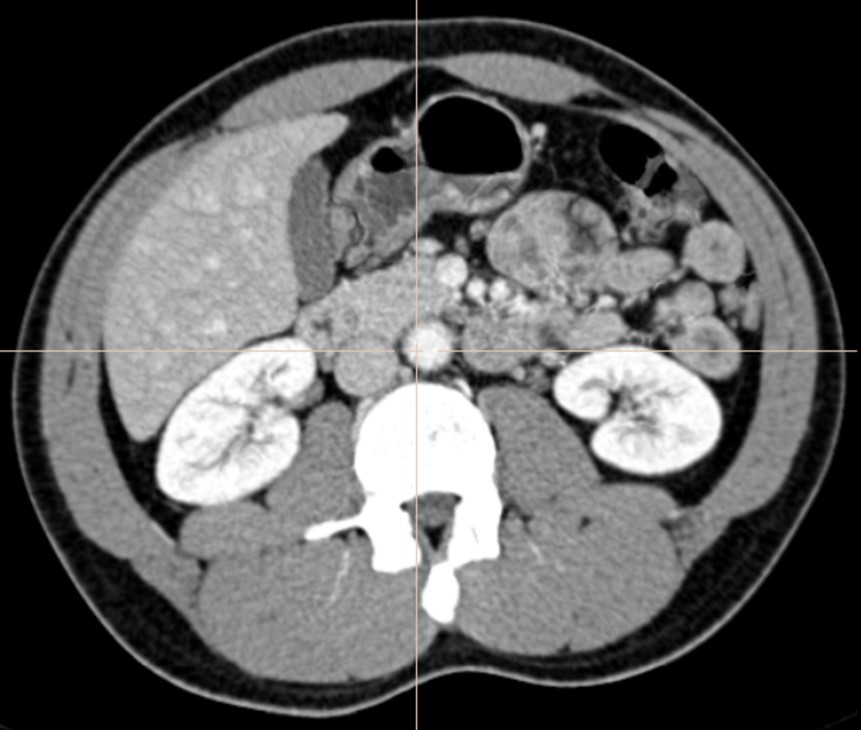}
\includegraphics[width=0.365\textwidth]{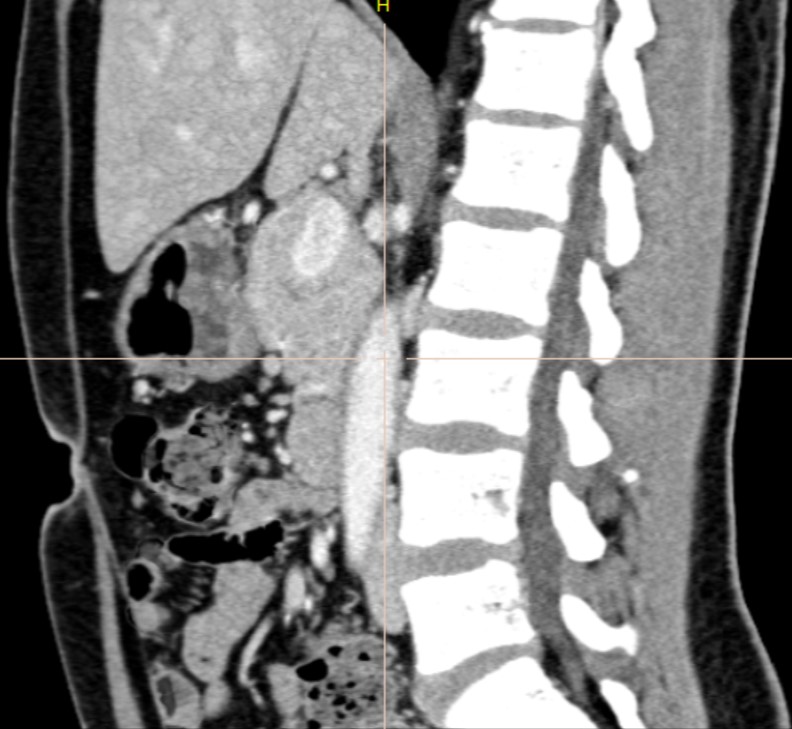}
\includegraphics[width=0.4\textwidth]{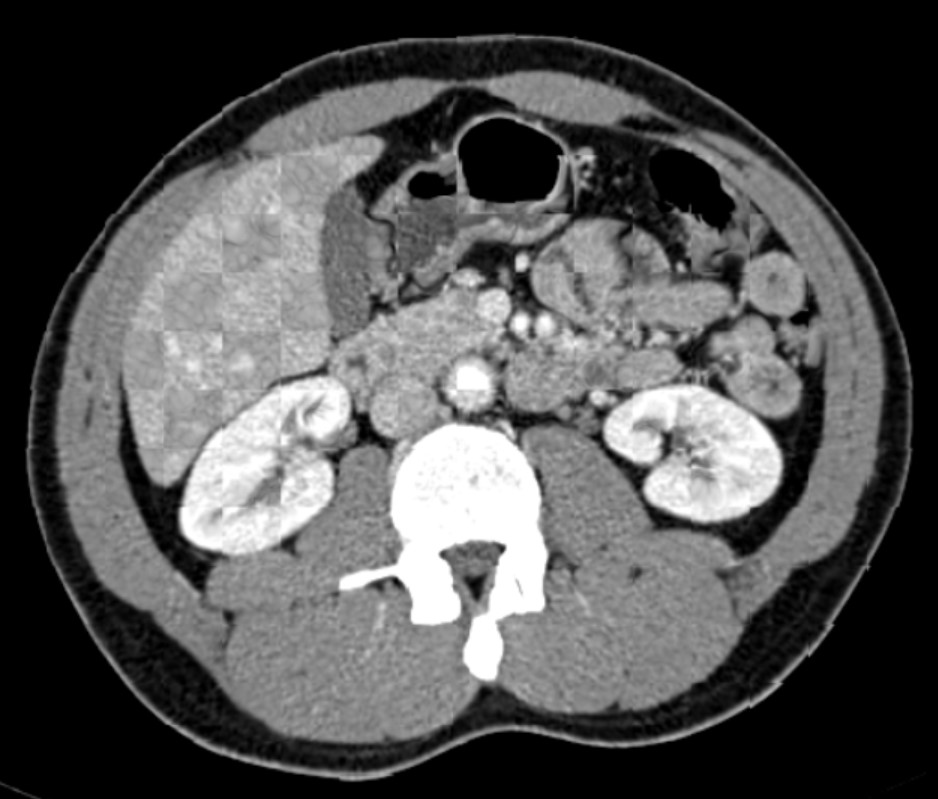}
\includegraphics[width=0.365\textwidth]{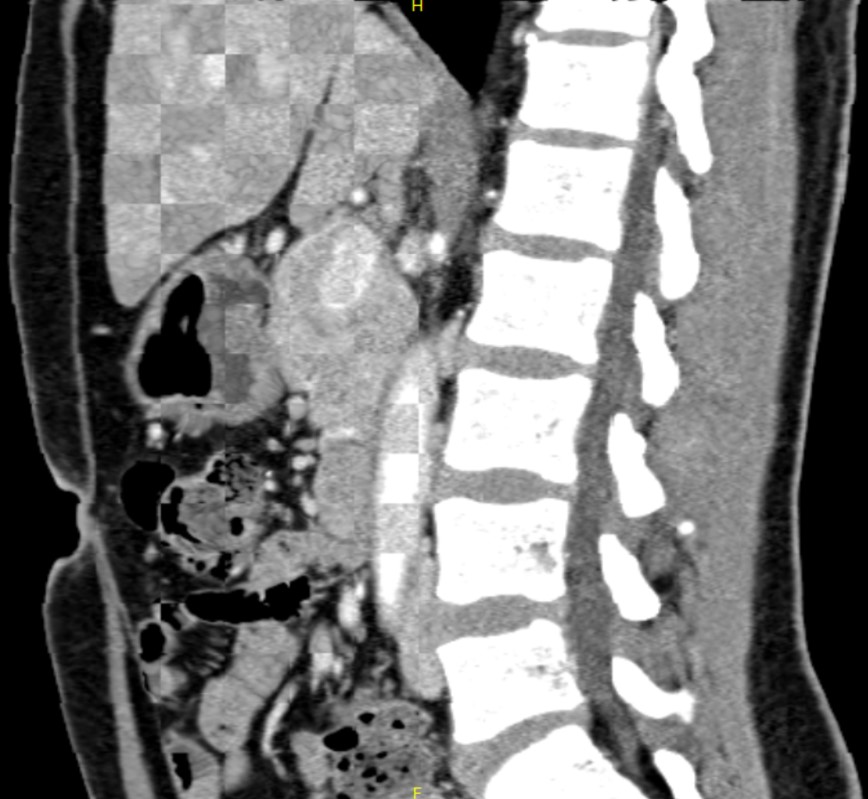}
\caption{A registration example of our algorithm. The top row shows venous phase axial (left) and sagittal (right) images, and the second row images are $50\%$ overlays of the original arterial phase to the venous phase. The third row images are the $50\%$ overlay of registered arterial phase image to the original venous image, and the fourth row images show checkerboard display (patch size: $16 \times 16$) of the result. All the images are visualized with the same window level and width.}
\end{figure*}

\section{Conclusion}
\label{sec:conclusion}

Despite the recent success of DNNs, especially in detection and segmentation in medical images, internal model is black-box type which is not interpretable nor controllable. Generative models are important to understand the underlying models and they can provide internal stage of the deformation. In this paper, we combined direction-dependent motion and local-self similarity to find the multi-phase deformable registration, especially abdominal region using conventional deformable registration approach. By experiments, our approach outperforms existing algorithms with efficient computation time which can be acceptable in various applications. Multi-modal images with larger motion for intra-subject registration can considered for the further study.

\bibliography{library}

\bibliographystyle{abbrv}

\end{document}